\documentclass{JHEP3}
\usepackage{amsmath}
\input epsf
\usepackage{epsfig}
\usepackage{amssymb}
\usepackage[active]{srcltx}

\newcommand{\bea}{\begin{eqnarray}}
\newcommand{\eea}{\end{eqnarray}}
\newcommand{\bean}{\begin{eqnarray*}}
\newcommand{\eean}{\end{eqnarray*}}
\newcommand{\nn}{\nonumber \\}

\def\W #1{\widetilde{#1}}
\def\WH #1{\widehat{#1}}

\def\braket#1{\left\langle #1 \right\rangle}

\def\ket#1{\left| #1\right\rangle}
\def\gb #1{ \left\langle #1 \right]}
\def\tgb #1{ \left[ #1 \right\rangle}

\def\a{{\alpha}}

\def\b{{\beta}}

\def\la{\lambda}

\def\vev{\braket}
\def\tgb #1{ \left[ #1 \right\rangle}
\def\bket#1{\left| #1\right]}
\def\bvev#1{\left[ #1 \right]}
\def\Spaa{\vev}
\def\Spbb{\bvev}
\def\Spab{\gb}
\def\Spba{\tgb}

\def\Label#1{\label{#1}%
  \smash{\hbox to0pt{\raise1ex\hbox{\tiny[#1]}\hss}}}

\title{Note on New KLT relations}
\author{Bo Feng$^{a,d}$, Song He$^{b,d}$, Rijun Huang$^{c,d}$, Yin
Jia$^{c,d}$\\
$^a$Center of Mathematical Science, Zhejiang University, Hangzhou,
China\\
$^b$\small Max-Planck-Institut f\"{u}r Gravitationsphysik,
Albert-Einstein-Institut, Golm, Germany\\
$^{c}$Zhejiang Institute of Modern Physics, Physics Department, Zhejiang University, Hangzhou, China\\
$^d$\small Kavli Institute for Theoretical Physics China, CAS,
Beijing 100190, China}
\date{Today}
\abstract{In this short note, we present two results about KLT
relations discussed in recent several papers. Our first result is
the re-derivation of Mason-Skinner MHV amplitude by applying the
$S_{n-3}$ permutation symmetric KLT relations directly to MHV
amplitude. Our second result is the equivalence proof of the newly
discovered $S_{n-2}$ permutation symmetric  KLT relations and the
well-known $S_{n-3}$ permutation symmetric KLT relations. Although
both formulas have been shown to be correct by BCFW recursion
relations, our result is the first direct check using the
regularized definition of the new formula. }

\begin{document}

\section{Introduction}

S-matrix program~\cite{S-matrix} is a program to study properties of
quantum field theory based on some general principles, like the
Lorentz invariance, Locality, Causality, Gauge symmetry as well as
Analytic property. Because it does not use specific information like
Lagrangian, result obtained by this method is quite general. Also
exactly because its generality with very few assumptions, study
along this line is very challenging.

One of the most important recent progresses in S-matrix program is
the derivation of BCFW recursion relations in gauge
theories~\cite{Britto:2004ap,Britto:2005fq} and
gravity~\cite{BCFW-Gra}, which relies only on basic analytic
properties of tree amplitudes if there are no boundary
contributions\footnote{The boundary behavior is one important
subject to study. In~\cite{ArkaniHamed:2008yf}, background field
method has been applied to the study. In~\cite{boundary}, the
situation with nonzero boundary contributions has also been
discussed. It will be interesting to study the boundary behavior in
the frame of S-matrix program.}. Furthermore, in~\cite{Paolo:2007},
by assuming the applicability of BCFW recursion relations in gauge
theories and gravity, many well-known (but difficult to prove)
fundamental facts about S-matrix, such as non-Abelian structure for
gauge theory and all matters couple to gravity with same coupling
constant, have been re-derived from S-matrix viewpoint\footnote{Gauge
theory three-point amplitudes are uniquely determined by Poincare
symmetry, in~\cite{SHST} it has been proved that, through BCFW
recursion relations, any higher-point tree amplitudes can be
consistently constructed if and only if there exists a non-Abelian
gauge group.}.

Based on these developments, non-trivial relations among tree-level
color-ordered gauge theory amplitudes, including the recently
proposed Bern-Carrasco-Johansson(BCJ) relations~\cite{Bern:2008qj}
(see also some applications \cite{BCJstringproofs}), have been
proved using BCFW recursion relations in~\cite{Feng:2010my}, which
provided the first field-theoretical (S-matrix) proof of these
relations\footnote{The BCJ relations have also been proved in string
theory~\cite{BjerrumBohr:2009rd,Stieberger:2009hq}.}. Using similar
ideas for gravity, new forms of Kawai-Lewellen-Tye(KLT) type
relations~\cite{Kawai:1985xq} (for a good review, see
\cite{Bern:2002kj}), which express gravity tree amplitudes as square
of gauge theory amplitudes, have been found and proved
in~\cite{BjerrumBohr:2010ta,BjerrumBohr:2010zb,Feng:2010br,BjerrumBohr:2010yc}.

There are two forms of KLT relations. The form with manifest
$S_{n-2}$ permutation symmetry is proposed and proved
in~\cite{BjerrumBohr:2010ta}. It is mostly suitable for a BCFW(pure
S-matrix) proof, but needs regularization to be well-defined. The
most general expression of the minimally (manifest $S_{n-3}$
permutation) symmetric form, is proposed and proved
in~\cite{BjerrumBohr:2010yc}, which has included the well-known
ansatz for KLT relations conjectured in~\cite{Bern:1998sv} as a
special case. This $S_{n-3}$ symmetric form is most natural from
string perspective, as originally proposed and proved in string
theory~\cite{Kawai:1985xq}. Both $S_{n-2}$ and $S_{n-3}$ symmetric
forms have been generalized to ${\cal N}=8$ SUGRA case with similar
S-matrix proofs in~\cite{Feng:2010br}, which naturally produce new
identities among ${\cal N}=4$ SYM amplitudes, including all 'flipped
identities' for gluon amplitudes~\cite{BjerrumBohr:2010zb}(see
also~\cite{Tye:2010kg}). Through string theory or BCFW recursion
relation, the equivalence relation between  $S_{n-2}$ and $S_{n-3}$
symmetric forms has been established. However, both methods are
indirect, thus a direct algebraic manipulation is desired. As a
major result of this note, in the second part we will show that
there is a direct derivation from $S_{n-2}$ symmetric form to the
minimal, $S_{n-3}$ symmetric form.

Although KLT relations give graviton amplitudes in terms of gluon
amplitudes no matter what the helicity configuration is, there are
not much explicit expressions available for graviton amplitudes
unlike the case of gluon amplitudes. Among all helicity
configurations, one of them is exceptional, i.e.,  the so-called MHV
(maximally-helicity-violating) amplitudes. In Yang-Mills case, the
famous Parke-Taylor formula~\cite{Parke:1986gb} for gluon MHV tree
amplitudes is
astonishingly simple. 
On the other hand, the case for gravity amplitudes is much more
complicated, even in the MHV sector. Various explicit formulas of
MHV gravity amplitudes have been
proposed~\cite{Berends:1988zp,Bern:1998sv,Nair:2005iv,Bedford:2005yy,Elvang:2007sg,Mason:2008jy,Nguyen:2009jk},
which fall into two categories: those with manifest $S_{n-2}$
permutation symmetry, such as the formula given by
Elvang-Freedman~\cite{Elvang:2007sg}, and those with $S_{n-3}$
symmetry, such as the original BGK formula~\cite{Berends:1988zp} and
the equivalent Mason-Skinner formula~\cite{Mason:2008jy}. However,
most of these formulas have been derived from approaches other than
KLT relations, and it is non-trivial to show that they are
equivalent to each other\cite{Spradlin:2008bu}. In the following, we
will show that one particularly simple formula, the Mason-Skinner
formula, directly follows from the $S_{n-3}$ symmetric KLT
relations, given the Parke-Taylor formula for gauge theory MHV
amplitudes as the input. In addition, we will discuss the relation
between Elvang-Freedman formula and the $S_{n-2}$ symmetric form of
KLT relations. As a byproduct, we will obtain an infinite number of
new formulas for gravity MHV amplitudes. The equivalence of all
these formulas are ensured by our derivation of $S_{n-3}$ symmetric
form from $S_{n-2}$ symmetric form.

The outline of the note is the following. In section two we will
derive Mason-Skinner formula for MHV gravity amplitudes from the
recently proposed $S_{n-3}$ permutation symmetric form of KLT
relations. In section three, we will show the equivalence of
$S_{n-2}$ symmetric form and $S_{n-3}$ symmetric form of KLT
relations, and as an application, we derive from the $S_{n-2}$
symmetric form an infinite number of new formulas for MHV gravity
amplitudes, which are equivalent to BGK formula and Elvang-Freedman
formula. In the Appendix we give another regularization procedure
for $S_{n-2}$ symmetric KLT formula.

\section{From new KLT to Mason-Skinner MHV gravity amplitude}

As mentioned in the introduction, although we have had general KLT
relations and in principle all graviton amplitudes can be obtained
through results of gluon amplitudes, so far most explicit formulas
for general graviton amplitudes are constrained to MHV-graviton
amplitudes\footnote{There are also some results for NMHV amplitudes
and the general algorithm for ${\cal N}=8$
SUSY-Gravity\cite{Bianchi:2008pu,Drummond:2009ge}.}. All these
expressions of MHV-graviton amplitudes are also very different and
it takes efforts to show the equivalence among
them\cite{Spradlin:2008bu}. One of these expressions is given by
Mason and Skinner\footnote{We have written results in the QCD
convention, which is different from the twistor convention by
$\Spbb{~}\to -\Spbb{~}$.}\cite{Mason:2008jy} as follows
\bea {\cal
M}_{MS}^{MHV}=(-)^{n-3}\sum_{P(2,\ldots,n-2)}{A^{MHV}(1,2,\ldots,n)\over
\Spaa{1|n-1}\Spaa{n-1|n}\Spaa{n|1}}\prod_{k=2}^{n-2}{\Spba{k|P_{k+1}+\cdots+P_{n-1}|n}\over
\Spaa{k|n}}~,~~ \label{msmhv} \eea
where the sum is over all $S_{n-3}$ permutations of labels
$(2,...,n-2)$. In this section we will show that starting from
general KLT formula and applying it to the MHV case, one can get the
Mason and Skinner formula.

Since the Mason and Skinner formula is with the sum over $S_{n-3}$
permutations, it is natural to start from following $S_{n-3}$
permutation symmetric KLT formula
\cite{BjerrumBohr:2010zb,BjerrumBohr:2010yc}
\bea {\cal M}_{n}^{KLT}=(-)^{n+1}\sum_{\alpha, \beta\in
S_{n-3}}A(1,\alpha,n-1,n){\cal
S}[\beta|\alpha]_{P_1}\tilde{A}(n,\beta,1,n-1)~,~~~\label{kltmhv}\eea
where the function ${\cal S}$ is defined as
\cite{BjerrumBohr:2010ta,BjerrumBohr:2010zb,BjerrumBohr:2010yc}
\bea {\cal S}[i_1,...,i_k|j_1,j_2,...,j_k]_{P_1} & = & \prod_{t=1}^k
(s_{i_t 1}+\sum_{q>t}^k \theta(i_t,i_q) s_{i_t
i_q})~~~\label{S-def}\eea
with $\theta(i_t, i_q)$ to be  zero when  pair $(i_t,i_q)$ has same
ordering at both sets ${\cal I},{\cal J}$ and otherwise, to be  one. Here $s_{ij}=(P_i+P_j)^2=2P_i\cdot P_j$.

Function ${\cal S}$ defined above has some properties which will be
useful for our discussions. The first one is the reversed property
\bea {\cal S}[i_1,...,i_k|j_1,j_2,...,j_k]_{P_1}={\cal
S}[j_k,...,j_1|i_k,..,i_1]_{P_1}~.~~~\label{S-rev}\eea
The second one is about the sum over permutations. To illustrate
this property, firstly we observe that
\bea
P_{ij}(S[\beta|\alpha]_{P_1})=S[P_{ij}(\beta)|P_{ij}(\alpha)]_{P_1}~,~~~\eea
where $P_{ij}$ is the permutation of label $i$ and label $j$ while
all other labels unchanged. Using this property we have
\bea
\sum_{\beta}S[\beta|P_{ij}(\alpha)]=\sum_{\beta}P_{ij}(S[P_{ij}(\beta)|\alpha])=
P_{ij}(\sum_{\beta}S[P_{ij}(\beta)|\alpha])=P_{ij}(\sum_{\beta}S[\beta|\alpha])~,~~~\eea
where at the third equal sign we have used the factor that sum over
all permutations $\sum_\b$ is commutative with particular
permutation $P_{ij}$. Then we have our second property
\bea \sum_{\alpha \beta} F(\b) S[\beta|\alpha]
G(\a)=\sum_{P(2,\ldots,n-2 )}(\sum_{\beta}
F(\b)S[\beta|2,\ldots,n-2] G(\{2,3,...,n-2\}))~,~~\label{pro-2}\eea
where $G(\a)$ is a general function. This property states that
although corresponding terms in both sides of (\ref{pro-2}) are
different under given permutations $\a,\b$, their summation is
equivalent.

After stating these useful properties of function $\cal{S}$, let us
continue our demonstration. Substituting (\ref{pro-2}) to formula
(\ref{kltmhv}) we get
\bea {\cal
M}_n^{KLT-MHV}=(-)^{n+1}\sum_{P(2,\ldots,n-2)}A^{MHV}(1,2,\ldots,n-1,n)
\sum_{\beta}S[\beta|2,\ldots,n-2]\tilde{A}^{MHV}(n,\beta,1,n-1)~.~~~\nonumber\\\label{kltmhv2}\eea
After comparing (\ref{msmhv}) with (\ref{kltmhv2}), we see that we
need to prove the following identity
\bea &&{1\over
\Spaa{1|n-1}\Spaa{n-1|n}\Spaa{n|1}}\prod_{k=2}^{n-2}{\Spba{k|P_{k+1}+\cdots+P_{n-1}|n}\over
\Spaa{k|n}}\nn &=&
\sum_{\beta}S[\beta|2,3,\ldots,n-2]\tilde{A}^{MHV}(n,\beta,1,n-1)~.~~~\label{peq}\eea
Note that label $(n-2)$ at the right hand part of function ${\cal
S}$ is at the last position, and in this case we can divide
permutations $\b\in S_{n-3}$ into groups of permutations $\gamma\in
S_{n-4}$ plus label $(n-2)$ inserted at all possible positions in
sequence fixed by $\gamma$. Using this observation we can write down
\bea &&\sum_{\beta}{\cal
S}[\beta|2,3,\ldots,n-2]_{P_1}A(n,\beta,1,n-1)\nn &=&\sum_{\gamma
\in P(2,\ldots,n-3)}\left(\sum_{\sigma\in OP(\{n-2\}\cup\{\gamma\})}{\cal
S}[\sigma|2,\ldots,n-2]_{P_1}A(n,\sigma,1,n-1)\right)\nn &=&
\sum_{\gamma}{\cal S}[\gamma|2,\ldots,n-3]_{P_1}
s_{n-2,n-1}A(n-2,n,\gamma,1,n-1)\nn &=& s_{n-2,n-1}\sum_{\gamma}{\cal
S}[\gamma|2,\ldots,n-3]_{P_1}A(n-2,n,\gamma,1,n-1)~,\label{genp}\eea
where $"OP"$ stands for "Ordered Permutation" and
$OP(\{\a\}\cup\{\beta\})$ means all possible permutations which
preserve elements' order of  both sets. Here, because the set
$\{n-2\}$ has only one element, $OP(\{n-2\}\cup\{\gamma\})$ just
indicates all possible $(n-2)$ insertions  into $\{\gamma\}$. Note
also that at the second line, the function ${\cal S}$ has $(n-3)$
labels while at the third line, only $(n-4)$ labels are left. In the
above derivation from the second line to the third line, we have
used
\bea  \sum_{\sigma\in OP(\{n-2\}\cup\{\gamma\})}{\cal
S}[\sigma|2,\ldots,n-2]_{P_1}A(n,\sigma,1,n-1)={\cal S}[\gamma|2,\ldots,n-3]_{P_1}
s_{n-2,n-1}A(n-2,n,\gamma,1,n-1)~,\nn\eea
which directly follows from the level one BCJ
relation(see~\cite{Feng:2010my}). As an example let us show the
details of calculation of $n=5$ case, which is
\bea &&\sum_{\sigma\in OP(\{3\}\cup\{2\})}{\cal
S}[\sigma|2,3]_{P_1}A(5,\sigma,1,4)\nn
&=& {\cal
S}[2,3|2,3]_{P_1}A(5,2,3,1,4)+{\cal
S}[3,2|2,3]_{P_1}A(5,3,2,1,4)\nn
&=& s_{21}\left(s_{31}A(5,2,3,1,4)+(s_{31}+s_{32})A(5,3,2,1,4)\right)\nn
&=& {\cal S} [2|2]_{P_1}s_{34}A(3,5,2,1,4)~,\eea
where in the fourth line, we have used ${\cal S} [2|2]_{P_1}=s_{21}$, the five-point level one BCJ relation
\bea s_{31}A(5,2,3,1,4)+(s_{31}+s_{32})A(5,3,2,1,4)+(s_{31}+s_{32}+s_{35})A(3,5,2,1,4)=0~,\eea
and the momentum conservation $s_{31}+s_{32}+s_{35}=-s_{34}$.

It is easy to see that, in the higher point case, $\sum_{\sigma\in
OP(\{n-2\}\cup\{\gamma\})}{\cal S}[\sigma|2,\ldots,n-2]_{P_1}$ can
always be written as a common factor ${\cal
S}[\gamma|2,\ldots,n-3]_{P_1}$ multiplying the corresponding
coefficients of level one BCJ relation, which we denote as
$f_1(n-2)$ in the next section, since it is irrelevant with other
elements except $(n-2)$.

We want to continue our simplification from the second line of
(\ref{genp}) to the third line. However, it seems not possible to do
any more simplification with the form given in (\ref{genp}) for
general amplitudes. But when dealing with MHV amplitudes, there is
"inverse soft factor"\cite{ArkaniHamed:2009dn} which relates
$(n-1)$-point MHV amplitude to $n$-point MHV amplitude as follows
(it can be easily seen from Parke-Taylor
formula~\cite{Parke:1986gb})
\bea
A^{MHV}(n-1,n-2,n,\gamma,1)={\Spaa{n-1|n}\over\Spaa{n-1|n-2}\Spaa{n-2|n}}
A^{MHV}(\W {n-1},\W n,\gamma,1,)~,~~~\label{soft} \eea
where in order to preserve the momentum conservation, i.e., $P_{\W
{n-1}}+P_{\W n}=P_{n-1}+P_{n-2}+P_n$, spinor components have been
modified as
\bea |\widetilde{n-1}]& = & {|P_{n-2}+P_{n-1}|n>\over
\Spaa{n-1|n}}~,~~~~|\widetilde{n-1}>=|n-1>~,~~~~\nn |\tilde{n}]& =&
{|P_{n-2}+P_{n}|n-1>\over
\Spaa{n|n-1}}~,~~~~|\tilde{n}>=|n>~.~~~~\label{changing}\eea
For (\ref{soft}) to be true  we have assumed that the helicity of
label $(n-2)$ is positive. This choice can always be made for
graviton MHV amplitudes where we can fix, for example, label $1,n$
to be negative helicities. Also, only anti-spinor parts of momenta
$P_{\W {n-1}},P_{\W n}$ have been changed while the spinor parts are
untouched. This observation will be very useful for our later
manipulation.

With this in mind we can continue our demonstration by substituting
(\ref{soft}) into (\ref{genp}) and get
\bea &&\sum_{\beta\in S_{n-3}}{\cal
S}[\beta|2,3,\ldots,n-2]_{P_1}A_n^{MHV}(n,\beta,1,n-1)\nn &=&
s_{n-2,n-1}{\Spaa{n-1|n}\over\Spaa{n-1|n-2}\Spaa{n-2|n}}\sum_{\gamma\in
S_{n-4}} {\cal
S}[\gamma|2,\ldots,n-3]_{P_1}A_{n-1}^{MHV}(\tilde{n},\gamma,1,\widetilde{n-1})~,~~~
\eea
where summation in the second line is similar to the one in the
first line except that the sum changes from $S_{n-3}$ to $S_{n-4}$.
Then we can iterate the procedure like the one did in (\ref{genp})
and yield
\bea &&\sum_{\beta}S[\beta|2,3,\ldots,n-2]A^{MHV}(n,\beta,1,n-1)\nn
&=&
s_{n-2,n-1}{\Spaa{n-1|n}\over\Spaa{n-1|n-2}\Spaa{n-2|n}}s_{n-3,\widetilde{n-1}}
{\Spaa{n-1|n}\over\Spaa{n-1|n-3}\Spaa{n-3|n}}\times\nn
&&\sum_{\gamma'\in P(2,\ldots,n-4)}{\cal
S}[\gamma'|2,\ldots,n-4]A^{MHV}(\tilde{\tilde{n}},\gamma',1,\widetilde{\widetilde{n-1}})\nn
&=&\cdots\nn &=&
S[2|2]A^{MHV}(\tilde{n}^{(n-4)},2,1,\widetilde{n-1}^{(n-4)})\prod_{k=3}^{n-2}
s_{k,\widetilde{n-1}^{(n-2-k)}}{\Spaa{n-1|n}\over\Spaa{n-1|k}\Spaa{k|n}}~,~~~\label{mhvp}\eea
where the notation $\tilde{n}^{(i)}$ means that there are $i$-th
changing of momentum $P_n$. Using (\ref{changing}) it is easy to get
the anti-spinor part of  $\tilde{n}^{(i)}$
\bea
|\widetilde{n-1}^{(i)}]&=&{|P_{n-1-i}+P_{\widetilde{n-1}^{(i-1)}}|n>\over
\Spaa{n-1|n}}={|P_{n-1-i}+P_{n-i}+P_{\widetilde{n-1}^{(i-2)}}|n>\over
\Spaa{n-1|n}}\nn
&=&\cdots={|P_{n-1-i}+P_{n-i}+\cdots+P_{n-2}+P_{n-1}|n>\over
\Spaa{n-1|n}}~.~~~\label{anti-i-times}\eea
Putting (\ref{anti-i-times}) back to (\ref{mhvp}) we get
\bea
&&S[2|2]A^{MHV}(\tilde{n}^{(n-4)},2,1,\widetilde{n-1}^{(n-4)})\prod_{k=3}^{n-2}s_{k,\widetilde{n-1}^{(n-2-k)}}{\Spaa{n-1|n}\over\Spaa{n-1|k}\Spaa{k|n}}\nn
&=& {1\over \Spaa{1|n-1}\Spaa{n-1|n}\Spaa{n|1}}{-\Spba{2|1|n}\over
\Spaa{2|n}}\prod_{k=3}^{n-2}{\Spaa{n-1|n}[k|\widetilde{n-1}^{(n-2-k)}]\over\Spaa{k|n}}\nn
&=& {1\over \Spaa{1|n-1}\Spaa{n-1|n}\Spaa{n|1}}{-\Spba{2|1|n}\over
\Spaa{2|n}}\prod_{k=3}^{n-2}{\Spba{k|P_{k+1}+P_{k+2}+\cdots+P_{n-1}|n}\over
\Spaa{k|n}}\nn &=& {1\over
\Spaa{1|n-1}\Spaa{n-1|n}\Spaa{n|1}}\prod_{k=2}^{n-2}{\Spba{k|P_{k+1}+P_{k+2}+\cdots+P_{n-1}|n}\over
\Spaa{k|n}}~,~~~\eea
where in the fourth line, we have rewritten $-\Spba{2|1|n}$ as
$\Spba{2|P_3+P_4+\cdots+P_{n-1}|n}$ using momentum conservation.
This finishes the proof of (\ref{peq}) .

Before ending this section, there is one application of above
derivation we want to address. In
\cite{BjerrumBohr:2010ta,BjerrumBohr:2010zb,Feng:2010br,Tye:2010kg},
new quadratic vanishing identities have been found and using them,
one can reduce the independent helicity bases from $(n-3)!$ down
further. For example, if we chose $A$ to be non-MHV and $\W A$ to be
MHV, we have
\bea 0=(-)^{n+1}\sum_{\alpha, \beta\in
S_{n-3}}A^{non-MHV}(1,\alpha,n-1,n){\cal
S}[\beta|\alpha]_{P_1}\tilde{A}^{MHV}(n,\beta,1,n-1)~.~~~\label{vanishing}\eea
Since identity  (\ref{peq}) is true as long as $\W A$ is MHV,  we
obtain immediately
\bea 0=\sum_{\a\in S_{n-3}(2,..,n-2)}
{A^{non-MHV}(1,\{2,3,..,n-2\},n-1,n)\over
\Spaa{1|n-1}\Spaa{n-1|n}\Spaa{n|1}}\prod_{k=2}^{n-2}{\Spba{k|P_{k+1}+\cdots+P_{n-1}|n}\over
\Spaa{k|n}}\eea
for amplitudes $A$ which are not MHV-amplitudes. This result has
been presented in \cite{Feng:2010br} where many other identities can
be written down too.

\section{From $S_{n-2}$ KLT to $S_{n-3}$ KLT}

One important result of recent study of KLT relations is the
manifest $S_{n-2}$ permutation symmetric KLT formula presented in
\cite{BjerrumBohr:2010ta}
\bea {\cal M}_n^{new}=(-1)^n\sum_{\gamma, \beta\in
S_{n-2}}{\tilde{A}_n(n,\gamma,1){\cal S}[\gamma|\beta]_{P_1}
A_n(1,\beta,n)\over s_{123\ldots(n-1)}}~.\label{sn-2klt}\eea
Formula (\ref{sn-2klt}) is not intuitive seeing from the familiar
KLT relations presented in \cite{Bern:1998sv}, even with the help of
new discovered BCJ relations\cite{Bern:2008qj}. However, as shown in
\cite{BjerrumBohr:2010yc}, this formula is the consistent
requirement of the pure field understanding of $S_{n-3}$ permutation
symmetric KLT relation under the BCFW expansion and in fact, it is
found by this way. Comparing to the formula given in
\cite{Bern:1998sv}, formula (\ref{sn-2klt}) is much easy to prove
using BCFW recursion relations in field theory while its stringy
derivation is still missing. Although formulas (\ref{kltmhv}) and
(\ref{sn-2klt}) are equivalent seen from BCFW recursion relations,
in this section we will try to establish more direct relation
between them.

\subsection{The direct derivation}

As emphasized in \cite{BjerrumBohr:2010ta, BjerrumBohr:2010yc},
naively (\ref{sn-2klt}) seems to be ill-defined since
$s_{123\ldots(n-1)}$ vanishes on-shell. However, there is a specific
regularization under which (\ref{sn-2klt}) is a well-defined finite
expression. The regularization is given by following off-shell
continuation of momenta $p_1$ and $p_n$ with an arbitrary momentum
$q$\cite{BjerrumBohr:2010ta, BjerrumBohr:2010yc}
\bea p_1\rightarrow p_1-xq,~~p_n\rightarrow p_n+xq~.\eea
In order to have the on-shell condition for $p_1$, we need to impose
$p_1\cdot q=0$ and $q^2=0$, while $q\cdot p_n \neq 0$. Thus we have
$p_{\hat{1}}^2=0$ and $p_{\hat{n}}^2=s_{\hat{1}23\ldots(n-1)}\neq
0$. Then a more accurate definition of (\ref{sn-2klt}) is the
following limit
\bea {\cal M}_n^{new}=(-1)^n\lim_{x\rightarrow 0 }\sum_{\gamma,
\beta\in
 S_{n-2}}{\tilde{A}_n(\hat{n},\gamma,\hat{1})S[\gamma|\beta]_{\WH P_1}A_n(\hat{1},\beta,\hat{n})\over
s_{\hat{1}23\ldots(n-1)}}~,~\label{rsn-2klt}\eea
where we have used "$\hat{\phantom{1}}$" to remind us the off-shell
regularization scheme. Now the denominator becomes
\bea s_{\hat{1}23\ldots(n-1)}=p_{\hat{n}}^2=(p_n+xq)^2=x\cdot
s_{nq}\neq 0~,~~~\eea
which means when taking the limit we only need to consider the
linear coefficient of $x$ in the numerator.

One important observation of (\ref{rsn-2klt}) is that we only need
to regularize one kind of these two amplitudes, because in the
numerator either combination
$\sum_{\beta}S[\gamma|\beta]A_n(1,\beta,n)$ or
$\sum_{\gamma}\tilde{A}_n(n,\gamma,1)S[\gamma|\beta]$ vanishes due
to the level one BCJ relation. If we denote, after regularization,
one combination to be $f(x)$, the remaining amplitudes to be $g(x)$
and the denominator to be $h(x)$, then we have\footnote{We would
like to thank T. Sondergaard for discussions on this point.}
\bea \left\{ \begin{aligned}&\lim_{x\rightarrow 0 }f(x)=0~,~~~
\lim_{x\rightarrow 0 }h(x)=0~,~~~ \lim_{x\rightarrow 0 }{f(x)\over
h(x)}=\mbox{const~.}&\\& \lim_{x\rightarrow 0 }g(x)\neq 0&
\end{aligned} \right.\nn
\Longrightarrow \lim_{x\rightarrow 0 }{g(x)f(x)\over
h(x)}=g(0)\cdot\lim_{x\rightarrow 0 }{f(x)\over
h(x)}=\lim_{x\rightarrow 0 }g(0){f(x)\over h(x)}~,~~~\eea
which shows directly that only one kind of amplitudes is needed to
be regularized. Without loss of generality we choose to regularize
$A_n(1,\beta,n)$, which simplifies (\ref{rsn-2klt}) to
\bea {\cal M}_n^{new}=(-1)^n\lim_{x\rightarrow 0 }\sum_{\gamma
\beta}{\tilde{A}_n(n,\gamma,1)S[\gamma|\beta]_{\WH P_1}A_n(\WH
1,\beta,\WH{n})\over
s_{\hat{1}23\ldots(n-1)}}~.~~~\label{rsn-2klt2}\eea

We want to simplify (\ref{rsn-2klt2}) further. As we have seen in
previous section, the last label in the sequence given by $\b$,
which denoted by $\b_{n-2}$, possess a nice property. Under this
consideration we regroup the two summations over $\gamma ,\beta$ as
follows
\bea &&\sum_{\gamma, \beta\in S_{n-2}} \tilde{A}_n(n,\gamma,1){\cal
S}[\gamma|\beta]_{\WH P_1}A_n(\hat{1},\beta,\hat{n})\nn
&=&\sum_{\beta\in S_{n-2}}A_n(\hat{1},\beta,\hat{n})\sum_{\gamma\in
S_{n-2}}\tilde{A}_n(n,\gamma,1) {\cal
S}[\gamma|\beta_1,...,\beta_{n-3},\beta_{n-2}]_{\WH P_1}\nn
&=&\sum_{\beta}A_n(\hat{1},\beta,\hat{n})\sum_{\gamma(\b_{n-2})\in
S_{n-3}}\left[\sum_{\sigma \in OP(\{\gamma(\b_{n-2})\}\cup
\{\b_{n-2}\})}\tilde{A}_n(n,\sigma,1){\cal S}[\sigma|\beta_1,...,\beta_{n-3},\beta_{n-2}]_{\WH P_1}\right]
,\label{pr1}\eea
where we have divided the permutation sum $\gamma\in S_{n-2}$ into
the permutation sum $\gamma(\b_{n-2})\in
S_{n-3}$\footnote{$\gamma(\b_{n-2})$ means the element $\b_{n-2}$
has been excluded.} plus all possible insertions of $\b_{n-2}$.
With the fixed $\b$-ordering, we have
\bea {\cal S}[\sigma|\beta_1,...,\beta_{n-3},\beta_{n-2}]_{\WH
P_1}={\cal S}[\gamma(\b_{n-2})|\beta_1,...,\beta_{n-3}]_{\WH P_1}
f_{\hat{1}}(\b_{n-2})~,~~~\eea
where $f_{\hat{1}}(\b_{n-2})$ (mentioned before) is the kinematic factor provided by
element $\b_{n-2}$. In other words, the dependence of  insertion
positions of $\b_{n-2}$ is given completely by the factor
$f_{\hat{1}}(\b_{n-2})$. The dependence of deformed momentum
$\hat{1}$ inside factor $f_{\hat{1}}(\b_{n-2})$ is given by
$s_{\b_{n-2},\hat{1}}=s_{\b_{n-2},1}-x s_{\b_{n-2},q}$, thus we have
\bea f_{\hat{1}}(\b_{n-2})= f_{1}(\b_{n-2})-x s_{\b_{n-2}, q}~.\eea
As mentioned before, the key point is that
\bea
\sum_{\sigma \in OP(\{\gamma(\b_{n-2})\}\cup
\{\b_{n-2}\})}\tilde{A}_n(n,\sigma,1)f_{1}(\b_{n-2})=0\eea
by level-one BCJ relation, since $\W A$ are un-deformed amplitudes.
Putting all together we finally have
\bea &&\sum_{\gamma, \beta\in S_{n-2}} \tilde{A}_n(n,\gamma,1){\cal
S}[\gamma|\beta]_{\WH P_1}A_n(\hat{1},\beta,\hat{n})\nn
&=&\sum_{\beta}A_n(\hat{1},\beta,\hat{n})\sum_{\gamma(\b_{n-2})\in
S_{n-3}}(-x
s_{\b_{n-2}, q}){\cal
S}[\gamma(\b_{n-2})|\beta_1,...,\beta_{n-3}]_{\WH P_1}\left[\sum_{\sigma \in OP(\{\gamma(\b_{n-2})\}\cup
\{\b_{n-2}\})}\tilde{A}_n(n,\sigma,1)\right]\nn
&=&\sum_{\beta}A_n(\hat{1},\beta,\hat{n})\sum_{\gamma(\b_{n-2})\in
S_{n-3}}(-\tilde{A}_n(n,\gamma(\b_{n-2}),1,\b_{n-2})  (-x
s_{\b_{n-2}, q}){\cal
S}[\gamma(\b_{n-2})|\beta_1,...,\beta_{n-3}]_{\WH
P_1}~,~~~\label{pr1-2}\eea
%
where we have used $U(1)$-decoupling relation for label $\b_{n-2}$
in the third line, i.e.,
\bea \sum_{\sigma \in OP(\{\gamma(\b_{n-2})\}\cup
\{\b_{n-2}\})}\tilde{A}_n(n,\sigma,1)=
-\tilde{A}_n(n,\gamma(\b_{n-2}),1,\b_{n-2})~.\eea

With expression (\ref{pr1-2})  we can take the limit
\bea {\cal M}_n^{new} & = & (-1)^n\lim_{x\rightarrow 0 }\sum_{\gamma
\beta}{\tilde{A}_n(n,\gamma,1)S[\gamma|\beta]_{\WH
P_1}A_n(\hat{1},\beta,\hat{n})\over s_{\hat{1}23\ldots(n-1)}}\nn
& = &(-1)^n\lim_{x\rightarrow 0 }
\frac{\sum_{\beta}A_n(\hat{1},\beta,\hat{n})s_{\b_{n-2}
q}\sum_{\gamma(\b_{n-2})\in
S_{n-3}}\tilde{A}_n(n,\gamma(\b_{n-2}),1,\b_{n-2}) {\cal
S}[\gamma(\b_{n-2})|\beta_1,...,\beta_{n-3}]_{\WH P_1} }{s_{nq}}\nn
& = &(-1)^n \frac{\sum_{\beta}A_n(1,\beta,n)s_{\b_{n-2}
q}\sum_{\gamma(\b_{n-2})\in
S_{n-3}}\tilde{A}_n(n,\gamma(\b_{n-2}),1,\b_{n-2}) {\cal
S}[\gamma(\b_{n-2})|\beta_1,...,\beta_{n-3}]_{P_1}
}{s_{nq}}~,~~\nonumber\\\label{Middle-new}\eea
where in the last step we have taken the $x\to 0$ limit so momenta
$p_1, p_n$ in $A$ are the un-deformed ones. In order to continue
further, we write the sum $\sum_{\b\in
S_{n-2}}=\sum_{\b_{n-2}=2}^{n-1}\sum_{\b(\b_{n-2})\in S_{n-3}}$ and
get
\bea {\cal M}_n^{new} & = &  -\sum_{\b_{n-2}=2}^{n-1}
{s_{\b_{n-2}q}\over s_{nq}} T_n(1,\b_{n-2},n)~,~~~\label{M-aux}\eea
with
\bea  &&T_n(1,\b_{n-2},n)\nn & = &
(-)^{n+1}\sum_{\b(\b_{n-2}),\gamma(\b_{n-2})\in S_{n-3}}A_n(1,\beta(\b_{n-2}),\b_{n-2},n){\cal
S}[\gamma(\b_{n-2})|\b(\b_{n-2})]_{P_1}\tilde{A}_n(n,\gamma(\b_{n-2}),1,\b_{n-2})~.\nn\eea
It is straightforward to see that $T_n(1,\b_{n-2},n)$ is nothing but
the graviton amplitude expression given in (\ref{kltmhv}) with
labels $1,n,\b_{n-2}$ fixed. Then if using the total symmetric
property of graviton amplitudes, we obtain immediately
\bea {\cal M}_n^{new} & = &  -\sum_{\b_{n-2}=2}^{n-1}
{s_{\b_{n-2}q}\over s_{nq}} T_n(1,\b_{n-2},n)= -{\cal
M}^{KLT}_n\sum_{\b_{n-2}=2}^{n-1} {s_{\b_{n-2}}\over s_{nq}}={\cal
M}^{KLT}_n ~,~~~~\label{M-proof}\eea
where in the last step we have used the momentum conservation and
$s_{1q}=0$.

There is one more thing we want to discuss before ending this part.
In our proof, in order to show that the new KLT formula with
manifest $S_{n-2}$ permutation symmetry is equivalent to the old KLT
formula with manifest $S_{n-3}$ permutation symmetry, we have used
the total symmetric property of old KLT formula, or at least the
$S_{n-2}$ permutation symmetry. This total symmetric property can be
seen from string theory, however it is not so obvious from field
theory. To show the property is true in field theory, one way is to
do algebraic manipulations using BCJ relations. However,
 with a few examples, it can be seen that such calculations are very
complicated with the increasing number of gravitons.

There is an indirect way to prove the total symmetric property of
KLT relations. The key point is to adapt induction method through
BCFW recursion relations. The three-point amplitudes are obviously
total symmetric by Lorentz symmetry and spin. Since graviton
amplitudes can be calculated by BCFW recursion relations, we can
build up higher point amplitudes from lower point amplitudes, which
have been assumed to be symmetric. Since all different KLT
expressions give same physical quantity, they must be equivalent to
each other, thus the total symmetric property is obtained. This idea
has already been used in \cite{BjerrumBohr:2010yc}.

\subsection{Application}

One obvious consequence of our proof is that if we do not use the
symmetry argument to pull out $T$ in (\ref{M-aux}), we will have a
new KLT formula with manifest $S_{n-2}$ permutation symmetry like
(\ref{sn-2klt}), but without the singular denominator. This formula
depends on an arbitrary auxiliary momentum $q$ as long as $q\cdot
p_1=0$. Applying (\ref{peq}) to (\ref{M-aux}), with some
manipulations we obtain
\bea {\cal M}_n^{new-MHV} & = &\sum_{\beta\in S_{n-2}} {\Spaa{n|1}
\Spaa{n-1|n-2} s_{(n-1)q}\over \Spaa{1|n-1}\Spaa{n|n-2} s_{nq}}
F(1,\{2,..,n-1\},n)~,~~~\label{Maux-MHV}\eea
where we have defined the following function
\bea F(1,2,..,n) & = & A(1,2,..,n) {  \Spaa{n|n-2}\over \Spaa{1|n}^2
\Spaa{n|n-1}\Spaa{n-1|n-2}} \prod_{s=2}^{n-2} {
\Spab{n|K_{(n-1)s}|s}\over \Spaa{n|s}}~~~\label{F-def}\eea
with $K_{(n-1)s}=p_{n-1}+p_{n-2}+...+p_s$. If we continue algebraic
manipulation like one has done from (\ref{M-aux}) to (\ref{M-proof})
we obtain
\bea {\cal M}_n^{BGK} & = &\sum_{\beta\in S_{n-3}} {\Spaa{1|n}
\Spaa{n-1|n-2} \over \Spaa{1|n-1}\Spaa{n|n-2} }
F(1,\{2,..,n-2\},n-1,n)~,~~~\label{Maux-BGK}\eea
which is nothing but the BGK expression \cite{Berends:1988zp}
rewritten  by Elvang and Freedman in \cite{Elvang:2007sg}. Using the
same function $F$, in \cite{Elvang:2007sg} a manifest $S_{n-2}$
permutation symmetric MHV amplitude is given by\footnote{ Using the
bonus relation\cite{Spradlin:2008bu}, it has been proved that
(\ref{EF-n-2}) is equivalent to (\ref{Maux-BGK}). }
\bea {\cal M}^{EF}_n & = & \sum_{\a\in S_{n-2}} F(1,\a\{2,3,...,
n-1\},n)~. ~~~\label{EF-n-2}\eea
Thus it is interesting to discuss the relation between
(\ref{Maux-MHV}) and (\ref{EF-n-2}).

In order to do so, we can simplify (\ref{Maux-MHV}) by taking
$q=\ket{1}\bket{q}$ so that $q\cdot p_1=0$, then we obtain
\bea {\cal M}^{MHV}_n & = &-\sum_{\beta\in S_{n-2}} { \Spaa{n-1|n-2}
\Spbb{n-1|q}\over \Spaa{n|n-2} \Spbb{n|q}}
F(1,\{2,..,n-1\},n)~.~~~\label{MHV-1}\eea
Formula (\ref{MHV-1}) is different from (\ref{EF-n-2}) and
(\ref{Maux-BGK}), but it can be  checked that all of them are
equivalent to each other  by BCFW recursion relations. A few
examples may be useful to demonstrate the relation between
(\ref{MHV-1}) and (\ref{EF-n-2}).  The case $n=3$ are simply $ -
{\Spab{1|2|q}\over \Spab{1|3|q}} F(1,2,3)=F(1,2,3)$ by momentum
conservation. For $n=4$ we have
\bean & & -F(1,2,3,4){\Spab{2|3|q}\over \Spab{2|4|q}}-F(1,3,2,4)
{\Spab{3|2|q}\over \Spab{3|4|q}}~. \eean
We can take a special case that $q=k_2$, then the second term is
zero and we obtain $-F(1,2,3,4){\Spab{2|3|2}\over \Spab{2|4|2}}$.
The BGK formula (\ref{Maux-BGK}) is $F(1,2,3,4)
{\Spaa{1|4}\Spaa{3|2}\over \Spaa{1|3}\Spaa{4|2}}$. In order to show
above two results are consistent we check the following expression
\bean  {{\Spab{2|3|2}\over \Spab{2|4|2}}\over
{\Spaa{1|4}\Spaa{3|2}\over \Spaa{1|3}\Spaa{4|2}}}=
{\Spab{1|4|2}\over \Spab{1|3|2}}=-1~,~~~\eean
which is true by momentum conservation.

We can learn from the different expressions (\ref{MHV-1}) and
(\ref{EF-n-2}) that the $S_{n-2}$ permutation symmetric form has
some redundancy, since the independent bases are $(n-3)!$ by BCJ
relations.

Although in this note, we are not able to change form (\ref{MHV-1})
to form (\ref{EF-n-2}) by direct algebraic manipulations, some
identities about function $F$ can be given by their equivalence.
When using $\Spab{1|n}$ BCFW-deformation
\bea \la_1(z)=\la_1+z \la_n,~~~\W\la_n(z)=\W \la_n-z \W\la_1~,\eea
$F(1,2,3,..,n)$  depends on $z$ only through factor ${1\over
\Spaa{1|2}}$ from $A(1,2,...,n)$, i.e., $F(1,2,...,n)$ contributes
to the pole $s_{12}(z)$ only. Now let us consider the residue given
by this pole from various MHV formulas. The formula (\ref{EF-n-2})
gives
\bea {\rm Res}({\cal M}^{EF}_{s_{12}})= {\Spaa{1|2}\over
\Spaa{n|2}}\sum_{\sigma\in S_{n-3}}
F(1,2,\sigma(3,...n-1),n)~,~~~\label{EF-s12}\eea
while the formula (\ref{MHV-1}) gives
\bea {\rm Res}({\cal M}^{new-MHV}_{s_{12}})= -{\Spaa{1|2}\over
\Spaa{n|2}}\sum_{\beta\in S_{n-3}} { \Spaa{n-1|n-2}
\Spbb{n-1|1}\over \Spaa{n|n-2} \Spbb{n|1}}
F(1,2,\{3,..,n-1\},n)~,~~~\label{our-s12-1}\eea
where we have taken $\bket{q}=\bket{1}$. The BGK formula gives
\bea {\rm Res}({\cal M}^{BGK-1}_{s_{12}})= {\Spaa{1|2}\over
\Spaa{n|2}}\sum_{P(3,..,n-2)} {\Spaa{2|n}\Spaa{n-1|n-2}\over
\Spaa{2|n-1}\Spaa{n|n-2}} F(1,2,\{3,...,n-2\},n-1,n)~,~~~\eea
and if we exchange $2\leftrightarrow (n-1)$ in BGK formula and take
the residue, we obtain
\bea {\rm Res}({\cal M}^{BGK-2}_{s_{12}}) & = &   \sum_{k=3}^{n-2}
{\Spaa{2|n-2}\Spaa{1|k}\over \Spaa{k|2}\Spaa{n|n-2}}\sum_\sigma
F(1,k,\sigma,2,n)~.~~~\eea
Since the residue is unique, above four expressions must be equal to
each other. Note that each expression has $(n-3)!$ terms, thus we
obtain relations between these $(n-3)!$ terms. This is consistent
with the new discovered relations given in
\cite{BjerrumBohr:2010ta,BjerrumBohr:2010zb,Feng:2010br,Tye:2010kg}.

\subsection*{Acknowledgements}

We would like to thank  discussions with N.E.J. Bjerrum-Bohr, P.H.
Damgaard and T. Sondergaard. BF, RJH, YJ are supported by fund from
Qiu-Shi, the Fundamental Research Funds for the Central Universities
with contract number 2009QNA3015, as well as Chinese NSF funding
under contract No.10875104. BF, SH, RJH, YJ thanks the organizers of
the program ``QFT, String Theory and Mathematical Physics'' at
KITPC, Beijing for hospitality while most parts of this work are
done.

\appendix

\section{The symmetry of graviton amplitude}

In section three, we have used the regularization procedure to show
the equivalence of new KLT formula given in
\cite{BjerrumBohr:2010ta} with the ones given in
\cite{Bern:1998sv,BjerrumBohr:2010zb,BjerrumBohr:2010yc}. There is
also a direct, but much more complicated way to check this. The good
point of this way is that we can see how the singular denominator
$s_{12...(n-1)}$ appears in the algebraic manipulation, thus in this
appendix we provide some details of this calculation. Before given
explicit example, let us write down following procedure of
calculations:

\begin{itemize}

\item {Step one:} Write down the expression (\ref{sn-2klt}).

\item {Step two:} Choose a minimal basis for  $A_n$ and $\widetilde{A}_n$.
These two basis (for $A$ and for $\W A$) can be different, but when
the choice has been made, it must be kept in following calculations.

\item {Step three:} Using  BCJ-relations to express all  remaining
amplitudes of $A$-type and $\widetilde{A}$-type in terms of the
chosen basis.

\item {Step four:}  Using  momentum conservation ($s_{in} = -\sum_{j=1}^{n-1} s_{ij}$)
to get rid of all $p_n$'s that might be in the BCJ-relations. In
other word, we have used $p_n=-\sum_{i=1}^{n-1} p_i$. But remember
we can not use $s_{1n}=s_{23..(n-1)}$.

\item {Step five:} Plugging  the $s_{in}$-free BCJ-relations into the expression
obtained from (\ref{sn-2klt}) and collecting corresponding
coefficients of each basis. Every coefficient must have factor
$s_{12...n-1}$ in numerator, thus we can cancel the same singular
factor in denominator.

\item {Step six:} After the pole is canceled we can go on-shell again and use
whatever known relations we want to reduce the  expression into the
familiar one, such as  (\ref{kltmhv}) etc.
\end{itemize}

The example we will demonstrate is the $n=5$ case
\bea (-)^5 M_5=& &  \W A(5,\a(2,3,4),1)\sum_{\a,\b} {\cal
S}[\a(2,3,4)|\b(2,3,4)] A(1,\b(2,3,4),5)\eea
Choosing $A(1,2,3,4,5)$ and $A(1,3,2,4,5)$ as a basis, other four
orderings are given as following\cite{Bern:2008qj}
\bea A(1,3,4,2,5) & = & { s_{12} A(1,2,3,4,5)+(s_{12}+s_{32})
A(1,3,2,4,5)\over s_{25}}\nn
A(1,4,3,2,5) & = & { s_{12} (s_{24}+s_{45}) A(1,2,3,4,5)- s_{13}
s_{24} A(1,3,2,4,5)\over s_{14} s_{25}}\nn
A(1,2,4,3,5) & = & { (s_{23}+s_{13})A(1,2,3,4,5)+s_{13}
A(1,3,2,4,5)\over s_{35}}\nn
A(1,4,2,3,5) & = & {-s_{12} s_{34}
A(1,2,3,4,5)-s_{13}(s_{14}+s_{24}) A(1,3,2,4,5)\over s_{14}
s_{35}}~~~\label{5-BCJ}\eea
It is worth to observe that the first and third one are the level
one BCJ relation, i.e., the denominator has only one $s_{ij}$, while
the second and fourth one are level two (with two $s_{ij}$ factors)
BCJ relations\footnote{Here we call the order of BCJ relations by
the number of denominator in formulas given in \cite{Bern:2008qj}.
It is worth to notice that while the level one BCJ relation has been
proved in \cite{BjerrumBohr:2009rd,Stieberger:2009hq,Feng:2010my},
higher order BCJ relations have not had a general proof although one
can explicit check it order by order recursively. It will be very
interesting to have a general proof.} . For general $n$, this
expansion needs to use up to level $(n-3)$ BCJ relations. Having the
result (\ref{5-BCJ}), we can calculate various terms by our rule (
do not forget to write, for example,
$s_{35}=-s_{31}-s_{32}-s_{34}$). For example, with
$\a(2,3,4)=(2,3,4)$ we have
\bean & & A(1,2,3,4,5) s_{21} s_{31} s_{41} + A(1,2,4,3,5) s_{21}
s_{41} (s_{31}+s_{43})\nn
& & + A(1,3,2,4,5) s_{31} (s_{21}+s_{23}) s_{41}+ A(1,4,3,2,5)
s_{41}(s_{31}+ s_{34}) (s_{21}+s_{23}+s_{24})\nn
& & + A(1,3,4,2,5) s_{31} s_{41} (s_{21}+s_{23}+s_{24})+
A(1,4,2,3,5) s_{41} (s_{21}+s_{24}) (s_{31}+s_{34})\\
& = & {s_{1234}\over s_{35}} (s_{31}+s_{34}) [ -s_{12}s_{34}
A(1,2,3,4,5)- s_{13} s_{24} A(1,3,2,4,5)] \eean
where the factor $s_{1234}$  appears in numerator.  Collecting all
six permutations together and getting rid of $s_{1234}$ we obtain
\bea & &   [ -s_{12}s_{34} A(1,2,3,4,5)- s_{13} s_{24} A(1,3,2,4,5)]
{(s_{31}+s_{34})\over s_{35}}\W A(2,3,4,1,5)\nn
&+ &  [ -s_{12}s_{34} A(1,2,3,4,5)- s_{13} s_{24} A(1,3,2,4,5)]
{s_{13}\over s_{35}} \W A(2,4,3,1,5) \nn
& + &  [ -s_{12}s_{34} A(1,2,3,4,5)- s_{13} s_{24} A(1,3,2,4,5)]
{(s_{21}+s_{24})\over s_{25}} \W A(3,2,4,1,5)\nn
& + &  [ -s_{12}s_{34} A(1,2,3,4,5)- s_{13} s_{24} A(1,3,2,4,5)]
 {s_{12}\over s_{25}}\W A(3,4,2,1,5)\nn
& + & \left[ -{s_{12} s_{34} s_{13}\over s_{35}}
A(1,2,3,4,5)-{s_{13}s_{24} (s_{21}+s_{23})\over s_{25}}
A(1,3,2,4,5)\right] \W A(4,2,3,1,5) \nn
& + & \left[ -{s_{12} s_{34} (s_{13}+s_{32})\over s_{35}}
A(1,2,3,4,5)-{s_{13}s_{24} s_{31}\over s_{25}} A(1,3,2,4,5)\right]
\W A(4,3,2,1,5)\eea
To continue, we add first four lines to get
\bea [ -s_{12}s_{34} A(1,2,3,4,5)- s_{13} s_{24} A(1,3,2,4,5)] (\W
A(3,2,4,1,5)+\W  A(2,3,4,1,5))~,\eea
and then add last two lines to get
\bea - s_{13} s_{24} A(1,3,2,4,5)\W A(2,4,3,1,5)-s_{12}s_{34}
A(1,2,3,4,5) \W A(3,4,2,1,5)~.\eea
Adding these two together we finally have
\bea & & s_{13} s_{24} A(1,3,2,4,5) \W A(2,4,1,3,5)+ s_{12}s_{34}
A(1,2,3,4,5)\W A(3,4,1,2,5)\nn
& = & -s_{13} s_{24} A(1,3,2,4,5) \W A(3,1,4,2,5)-s_{12} s_{34}
A(1,2,3,4,5) \W A(2,1,4,3,5)\eea
which is the familiar KLT relations.

From this example, it can be seen that the direct method is very
complicated  because we need to use various BCJ relations up to
level $(n-3)$ and to sum up various terms to obtain an overall
factor $s_{12..(n-1)}$. After got rid of $s_{12...(n-1)}$ from the
sum over $A$, we need to use BCJ relations again to sum over $\W A$.
Although case by case one can check, it is hard to observe the
general patterns to give a rigorous proof, thus it is better to use
our regularization method to give the proof as we did in section
three.



\begin{thebibliography}{999}

\bibitem{S-matrix} D.I. Olive, Phys. Rev. 135,B 745(1964); G.F.
Chew, "The Analytic S-Matrix: A Basis for Nuclear Democracy",
W.A.Benjamin, Inc., 1966; R.J. Eden, P.V. Landshoff, D.I. Olive,
J.C. Polkinghorne, "The Analytic S-Matrix", Cambridge University
Press, 1966.


\bibitem{Britto:2004ap}
  R.~Britto, F.~Cachazo and B.~Feng,
  Nucl.\ Phys.\  B {\bf 715}, 499 (2005)
  [arXiv:hep-th/0412308].

\bibitem{Britto:2005fq}
  R.~Britto, F.~Cachazo, B.~Feng and E.~Witten,
  Phys.\ Rev.\ Lett.\  {\bf 94}, 181602 (2005)
  [arXiv:hep-th/0501052].

\bibitem{BCFW-Gra}
  F.~Cachazo and P.~Svrcek,
  arXiv:hep-th/0502160.

  P.~Benincasa, C.~Boucher-Veronneau and F.~Cachazo,
  JHEP {\bf 0711}, 057 (2007)
  [arXiv:hep-th/0702032].

  J.~Bedford, A.~Brandhuber, B.~J.~Spence and G.~Travaglini,
  Nucl.\ Phys.\  B {\bf 721}, 98 (2005)
  [arXiv:hep-th/0502146].



\bibitem{ArkaniHamed:2008yf}
  N.~Arkani-Hamed and J.~Kaplan,
  JHEP {\bf 0804}, 076 (2008)
  [arXiv:0801.2385 [hep-th]].


\bibitem{boundary}
  B.~Feng, J.~Wang, Y.~Wang and Z.~Zhang,
  JHEP {\bf 1001}, 019 (2010)
  [arXiv:0911.0301 [hep-th]].

  B.~Feng and C.~Y.~Liu,
  arXiv:1004.1282 [hep-th].

\bibitem{Paolo:2007}
  Paolo.~Benincasa,   Freddy.~Cachazo,
  arXiv:0705.4305[hep-th].
\bibitem{SHST}
  S.~He and H.~b.~Zhang,
  JHEP {\bf 1007}, 015 (2010)
  [arXiv:0811.3210 [hep-th]].

  P.~C.~Schuster and N.~Toro,
  JHEP {\bf 0906}, 079 (2009)
  [arXiv:0811.3207 [hep-th]].

\bibitem{Bern:2008qj}
  Z.~Bern, J.~J.~M.~Carrasco and H.~Johansson,
  Phys.\ Rev.\  D {\bf 78}, 085011 (2008)
  [arXiv:0805.3993 [hep-ph]].

\bibitem{Feng:2010my}
  B.~Feng, R.~Huang and Y.~Jia,
  arXiv:1004.3417 [hep-th].


\bibitem{BjerrumBohr:2009rd}
  N.~E.~J.~Bjerrum-Bohr, P.~H.~Damgaard and P.~Vanhove,
  Phys.\ Rev.\ Lett.\  {\bf 103}, 161602 (2009)
  [arXiv:0907.1425 [hep-th]].


\bibitem{Stieberger:2009hq}
  S.~Stieberger,
  arXiv:0907.2211 [hep-th].

\bibitem{BCJstringproofs}
  H.~Tye and Y.~Zhang,
  arXiv:1003.1732 [hep-th].

  N.~E.~J.~Bjerrum-Bohr, P.~H.~Damgaard, T.~Sondergaard and P.~Vanhove,
  arXiv:1003.2403 [hep-th].





\bibitem{Kawai:1985xq}
  H.~Kawai, D.~C.~Lewellen and S.~H.~H.~Tye,
  Nucl.\ Phys.\  B {\bf 269}, 1 (1986).

\bibitem{Bern:2002kj}
  Z.~Bern,
  Living Rev.\ Rel.\  {\bf 5}, 5 (2002)
  [arXiv:gr-qc/0206071].



\bibitem{BjerrumBohr:2010ta}
  N.~E.~J.~Bjerrum-Bohr, P.~H.~Damgaard, B.~Feng and T.~Sondergaard,
  arXiv:1005.4367 [hep-th].

\bibitem{BjerrumBohr:2010zb}
  N.~E.~J.~Bjerrum-Bohr, P.~H.~Damgaard, B.~Feng and T.~Sondergaard,
  arXiv:1006.3214 [hep-th].

\bibitem{Feng:2010br}
  B.~Feng and S.~He,
  arXiv:1007.0055 [hep-th].

\bibitem{BjerrumBohr:2010yc}
  N.~E.~J.~Bjerrum-Bohr, P.~H.~Damgaard, B.~Feng and T.~Sondergaard,
  arXiv:1007.3111 [hep-th].

\bibitem{Bern:1998sv}
  Z.~Bern, L.~J.~Dixon, M.~Perelstein and J.~S.~Rozowsky,
  Nucl.\ Phys.\  B {\bf 546}, 423 (1999)
  [arXiv:hep-th/9811140].


\bibitem{Tye:2010kg}
 H.~Tye and Y.~Zhang,
  arXiv:1007.0597 [hep-th].

\bibitem{Parke:1986gb}
 S.~J.~Parke and T.~R.~Taylor,
 Phys.\ Rev.\ Lett.\ {\bf 56} (1986) 2459.

\bibitem{Berends:1988zp}
  F.~A.~Berends, W.~T.~Giele and H.~Kuijf,
  Phys.\ Lett.\  B {\bf 211}, 91 (1988).

\bibitem{Nair:2005iv}
  V.~P.~Nair,
  Phys.\ Rev.\  D {\bf 71}, 121701 (2005)
  [arXiv:hep-th/0501143].

\bibitem{Bedford:2005yy}
  J.~Bedford, A.~Brandhuber, B.~J.~Spence and G.~Travaglini,
  Nucl.\ Phys.\  B {\bf 721}, 98 (2005)
  [arXiv:hep-th/0502146].

\bibitem{Elvang:2007sg}
  H.~Elvang and D.~Z.~Freedman,
  JHEP {\bf 0805}, 096 (2008)
  [arXiv:0710.1270 [hep-th]].

\bibitem{Mason:2008jy}
  L.~Mason and D.~Skinner,
  arXiv:0808.3907 [hep-th].

\bibitem{Nguyen:2009jk}
  D.~Nguyen, M.~Spradlin, A.~Volovich and C.~Wen,
  arXiv:0907.2276 [hep-th].

\bibitem{Spradlin:2008bu}
  M.~Spradlin, A.~Volovich and C.~Wen,
  Phys.\ Lett.\  B {\bf 674}, 69 (2009)
  [arXiv:0812.4767 [hep-th]].


\bibitem{Bianchi:2008pu}
  M.~Bianchi, H.~Elvang and D.~Z.~Freedman,
  JHEP {\bf 0809}, 063 (2008)
  [arXiv:0805.0757 [hep-th]].

\bibitem{Drummond:2009ge}
  J.~M.~Drummond, M.~Spradlin, A.~Volovich and C.~Wen,
  Phys.\ Rev.\  D {\bf 79}, 105018 (2009)
  [arXiv:0901.2363 [hep-th]].

\bibitem{ArkaniHamed:2009dn}
  N.~Arkani-Hamed, F.~Cachazo, C.~Cheung and J.~Kaplan,
  JHEP {\bf 1003}, 020 (2010)
  [arXiv:0907.5418 [hep-th]].








\end{thebibliography}
\end{document}